\documentclass[aps,preprint,groupedaddress,showpacs]{revtex4}

\usepackage{amsmath}

\begin{document}


\title{Once again about beam-size or MD-effect\\ at colliding beams}
\author{G.L.~Kotkin and V.G.~Serbo}
\email{serbo@math.nsc.ru}
 \thanks{ This work is supported in part by INTAS
(code 00-00679), RFBR (code 00-02-17592) and by St. Petersburg
grant (code E00-3.3-146)}
 \affiliation{Novosibirsk State University, 630090, Novosibirsk, Russia}
 \date{December 6, 2002}

\begin{abstract}

For several processes at colliding beams, macroscopically large
impact parameters give an essential contribution to the standard
cross section. These impact parameters may be much larger than the
transverse sizes of the colliding bunches. In that case, the
standard calculations have to be essentially modify. The
corresponding formulae were given twenty years ago. In recent
paper of Baier and Katkov \cite{BK02} it was claimed that the
previous results about bremsstrahlung spectrum have to be revised
and an additional subtraction related to the coherent contribution
has to be done. This additional term has been calculated with the
result that it may be essential for the performed and future
experiments. In the present paper we analyze in detail the
coherent and incoherent contributions in the conditions,
considered in paper~\cite{BK02}. In contract to above claims, we
found out that under these conditions the coherent contribution is
completely negligible and, therefore, there is no need to revise
the previous results.

\end{abstract}


\maketitle

\section{INTRODUCTION}

\subsection{Beam-size or MD-effect}

The so called beam-size or MD-effect is a phenomenon discovered in
experiments \cite{Blinov82} at the MD-1 detector (the VEPP-4
accelerator with $e^+e^-$ colliding beams , Novosibirsk 1981). It
was found out that for ordinary bremsstrahlung, macroscopically
large impact parameters should be taken into consideration. These
impact parameters may be much larger than the transverse sizes of
the interacting particle bunches. In that case, the standard
calculations, which do not take into account this fact, will give
incorrect results. The detailed description of the MD-effect can
be found in review \cite{KSS}.

In $1980$--$1981$ a dedicated study of the process $e^+ e^-
\rightarrow e^+ e^- \gamma$ has been performed at the collider
VEPP-$4$ in Novosibirsk using the detector MD-1 for an energy of
the electron and positron beams $E_e=E_p = 1.8$ GeV and in a wide
interval of the photon energy $E_\gamma$ from $0.5$ MeV to
$E_\gamma \approx E_e$. It was observed \cite{Blinov82} that the
number of measured photons was smaller than that expected. The
deviation from the standard calculation reached $30 \%$ in the
region of small photon energies and vanished for large energies of
the photons. A.~Tikhonov \cite{Tikhonov82} pointed out that those
impact parameters $\varrho$, which give an essential contribution
to the standard cross section, reach values of $\varrho_m \sim 5$
cm whereas the transverse size of the bunch is $\sigma_\perp \sim
10^{-3}$ cm. The limitation of the impact parameters to values
$\varrho \lesssim \sigma_\perp$ is just the reason for the
decreasing number of observed photons.

The first calculations of this effect have been performed in Refs.
\cite{BKS} and \cite{BD} using different versions of
quasi--classical calculations in the region of large impact
parameters. Further experiments, including the measurement of the
radiation probability as function of the beam parameters,
supported the concept that the effect arises from the limitation
of the impact parameters. Later on, the effect of limited impact
parameters was taken into account when the single bremsstrahlung
was used for measuring the luminosity at the VEPP--$4$
collider~\cite{Blinov88} and at the LEP-I collider~\cite{Bini94}.
In the case of the VEPP--$4$ experiment~\cite{Blinov88}, it was
checked that the luminosities obtained using either this process
or using other reactions (such as the double bremsstrahlung
process $e^+e^- \rightarrow e^+ e^- \gamma \gamma$, where the
MD-effect is absent) agreed with each other.

A general scheme  to calculate the finite beam size effect had
been developed in paper~\cite{KPS85a} starting from the quantum
description of collisions as an interaction of wave packets
forming bunches. Since the effect under discussion is dominated by
small momentum transfer, the general formulae can be considerably
simplified. The corresponding approximate formulae were given.
They are obtained from an analysis of Feynman diagrams and it
allows to estimate the accuracy of approximation. In a second
step, the transverse motion of the particles in the beams can be
neglected. The less exact (but simpler) formulae, which are then
found, correspond to the results of Refs.~\cite{BKS} and
\cite{BD}. It has also been shown that similar effects have to be
expected for several other reactions such as bremsstrahlung for
colliding $ep$--beams~\cite{KPS85b}, \cite{KPSS88}, $e^+e^-$--
pair production in $e^\pm e$ and $\gamma e$
collisions~\cite{KPS85a}. The corresponding corrections to the
standard formulae are now included in programs for simulation of
events at linear colliders. The influence of MD-effect on
polarization had been considered in Ref.~\cite{KKSS89}.

The possibility to create high-energy colliding $\mu^+\mu^-$ beams
is now wildly discussed. For several processes at such colliders a
new type of beam-size effect will take place --- the so called
linear beam-size effect~\cite{KMS}. The calculation of this effect
had been performed by method developed for MD-effect
in~\cite{KPS85a}.

In 1995 the MD-effect was experimentally  observed at the
electron-proton collider HERA~\cite{Piot95} on the level predicted
in~\cite{KPSS88}.

It was realized in last years that  MD-effect in bremsstrahlung
plays important role for the problem of beam lifetime. At storage
rings TRISTAN and LEP-I, the process of a single bremsstrahlung
was the dominant mechanism for the particle losses in beams. If
electron loses more than $1\;\%$ of its energy, it leaves the
beam. Since  MD-effect reduced considerable the effective cross
section of this process, the calculated beam lifetime in these
storage rings was larger by about $25 \; \%$ for
TRISTAN~\cite{Funakoshi} and by about $40 \; \%$ for
LEP-I~\cite{Burkhard} (in accordance with the experimental data)
then without taken into account the MD-effect.

According to our calculations~\cite{KS02}, at B-factories PEP-II
and KEKB the MD-effect reduces beam losses due to
brems\-strah\-lung by about 20\%.

It is seen from this brief listing that the MD-effect is a
phenomenon interesting from the theoretical point of view and
important from the experimental point of view.

\subsection{Essence of the Baier-Katkov paper}

In recent paper~\cite{BK02}, previous results~\cite{BKS},
\cite{BD}, \cite{KPS85a} about bremsstrahlung spectrum had been
revised. It was claimed that an additional ``subtraction
associated with the extraction of pure fluctuation process'' has
to be done. The reason to perform such an subtraction explained as
follows: ``At the beam collision the momentum transfer may arise
due to interaction of the emitting particle with the opposite beam
as a whole (due to coherent interaction with average field of the
beam) and due to interaction with an individual particle of the
opposite beam. Here we consider {\it the incoherent} process only
(connected with the incoherent fluctuation of density) and so we
have to subtract the coherent contribution''. Analysis, performed
in paper~\cite{BK02}, results in the conclusion that this
additional ``subtraction term'' in the spectrum is not important
for the MD-1 experiment~\cite{Blinov82}, but it should be taken
into account in processing the HERA experiment~\cite{Piot95}; it
also may be important for the future experiments at linear
$e^+e^-$ colliders. It should be noted that in paper~\cite{BK02}
there is no derivation of the starting formulae, and all physical
reasons for such a subtraction were cited above. On the other
hand, in paper~\cite{BK02} there is a general remark that their
consideration was motivated by corresponding calculations for
bremsstrahlung of ultra-relativistic electrons on oriented
crystals.

In the present paper we analyze the coherent and incoherent
contributions in the conditions, considered in paper~\cite{BK02},
when the coherent length $l_{\rm coh}$ is much smaller than the
bunch length $l$ but much larger than the mean distance between
particles $a$, i.e. $a\ll l_{\rm coh} \ll l$. We derive
expressions for the coherent and incoherent contributions and show
that under these conditions the coherent contribution is
completely negligible and, therefore, there is no need to revise
the previous results. This conclusion is quite natural. A usual
bunch at colliders can be considered as a gaseous media with a
smooth particle distribution which has characteristic scales of
the order of bunch sizes. In particular, the average particle
density in such a bunch has the only scale in the longitudinal
direction --- the length of the bunch $l$. Therefore, the average
field of the bunch has the spectral components in the region of
frequancies $\omega=q_z c \sim c/l_{\rm coh} \lesssim c/l$ and
vanishes in the region of much higher frequencies considered here.
On the contrary, in the crystal case there is another scale
related to the size of the particle localization in the crystal
structure. In this case, the additional subtraction should be
taken into account for incoherent contribution. It seems that the
electron radiation on oriented crystals played a misleading role
for consideration of the MD-effect in~\cite{BK02}. To clarify a
question we present our calculations in full details.

\section{QUALITATIVE DESCRIPTION OF THE MD-EFFECT}

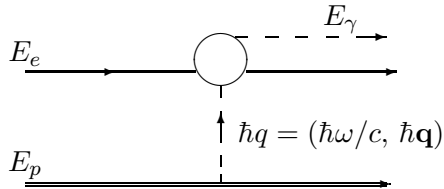
\begin{figure}[!htb]
  \centering
  \setlength{\unitlength}{1cm}
\unitlength=2.0mm \special{em:linewidth 0.4pt}
\linethickness{0.4pt}
\begin{picture}(26.00,15.00)

\put(1.00,1.80){\line(1,0){24.00}}
\put(1.00,1.60){\line(1,0){24.00}}
\put(25.2,1.70){\vector(1,0){0.10}}

\put(1.00,9.20){\vector(1,0){6.00}}
\put(14.00,10.00){\circle{3.40}} \put(7.00,9.2){\line(1,0){5.3}}
\put(15.70,9.20){\vector(1,0){10.00}}
\put(15.0,11.5){\line(1,0){0.8}} \put(17,11.5){\line(1,0){0.8}}
\put(19,11.5){\line(1,0){0.8}} \put(21,11.5){\line(1,0){0.8}}
\put(23,11.5){\vector(1,0){2.00}}

\put(14.00,4.50){\vector(0,1){2.00}}
\put(14.00,2.90){\line(0,1){0.71}}
\put(14.00,1.80){\line(0,1){0.51}}
\put(14.00,7.50){\line(0,1){0.7}}

\put(1.00,10.4){\makebox(0,0)[cc]{$E_e$}}
\put(1.00,3.00){\makebox(0,0)[cc]{$E_p$}}
\put(22.00,12.70){\makebox(0,0)[cc]{$E_\gamma$}}
\put(22.00,5.00){\makebox(0,0)[cc]{$\hbar q=(\hbar\omega/c,\,
\hbar{\bf q})$}}

\end{picture}
    \caption{Block diagram of radiation by the electron.}
 \label{fig:1}
  \end{figure}

\begin{figure}[!htb]
  \centering
\unitlength=2.00mm \special{em:linewidth 0.4pt}
\linethickness{0.4pt}
\begin{picture}(47.00,15.00)
\put(2.00,9.20){\vector(1,0){2.00}}
\put(10.00,10.00){\circle{3.40}} \put(3.00,9.20){\line(1,0){5.30}}
\put(11.70,9.20){\vector(1,0){6.00}}
\put(10.00,5.00){\vector(0,1){2.00}}
\put(10.00,3.00){\line(0,1){1.00}}
\put(10.00,7.80){\line(0,1){0.50}}
\
\
\put(11.00,11.50){\line(1,0){0.80}}
\put(13.00,11.50){\line(1,0){0.80}}
\put(15.00,11.50){\vector(1,0){2.00}}
\
\put(25.00,9.00){\vector(1,0){9.00}}
\put(38.00,9.00){\vector(1,0){9.00}}

\put(27.00,8.00){\line(0,1){1.00}}
\put(27.00,5.00){\vector(0,1){2.00}}
\put(27.00,3.00){\line(0,1){1.00}}

\put(32.00,9.00){\line(0,1){1.00}}
\put(32.00,11.00){\vector(0,1){2.00}}
\put(32.00,14.00){\line(0,1){1.00}}

\put(40.00,9.00){\line(0,1){1.00}}
\put(40.00,11.00){\vector(0,1){2.00}}
\put(40.00,14.00){\line(0,1){1.00}}

\put(45.00,8.00){\line(0,1){1.00}}
\put(45.00,5.00){\vector(0,1){2.00}}
\put(45.00,3.00){\line(0,1){1.00}}

\put(21.00,9.00){\makebox(0,0)[cc]{=}}
\put(36.00,9.00){\makebox(0,0)[cc]{+}}

\end{picture}
 \caption{Compton scattering of equivalent photon on the electron.}
 \label{fig:2}
\end{figure}
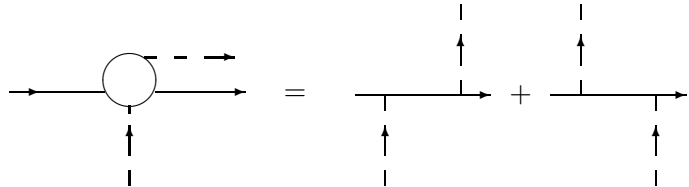

Qualitatively we describe the MD--effect using as an example the
$e p \rightarrow e p \gamma$ process\footnote{Below we use the
following notations: $N_e$ and $N_p$ are the numbers of electrons
and protons (positrons) in the bunches, $\sigma_z=l$ is the
longitudinal, $\sigma_x$ and $\sigma_y$ are the horizontal and
vertical transverse sizes of the proton (positron) bunch,
$\gamma_e=E_e/(m_ec^2)$, $\gamma_p=E_p/(m_pc^2)$ and $r_e=e^2/(m_e
c^2)$ is the classical electron radius.}. This reaction is defined
by the diagrams of Fig.~\ref{fig:1} which describe the radiation
of the photon by the electron (the contribution of the photon
radiation by the proton can be neglected). The large impact
parameters $\varrho \gtrsim \sigma_\perp$, where $\sigma_\perp$ is
the transverse beam size, correspond to small momentum transfer
$\hbar q_\perp \sim (\hbar / \varrho) \lesssim (\hbar /
\sigma_\perp)$. In this region, the given reaction can be
represented as a Compton scattering (Fig.~\ref{fig:2}) of the
equivalent photon, radiated by the proton, on the electron. The
equivalent photons with frequency $\omega$ form a ``disk'' of
radius $\varrho_m \sim \gamma_p c / \omega$ where $\gamma_p = E_p
/ (m_p c^2)$ is the Lorentz-factor of the proton. Indeed, the
electromagnetic field of the proton is $\gamma_p$--times
contracted in the direction of motion. Therefore, at distance
$\varrho$ from the axis of motion a characteristic longitudinal
length of a region occupied by the field can be estimated as
$\lambda \sim \varrho / \gamma_p$ which leads to the frequency
$\omega \sim c / \lambda \sim \gamma_p c / \varrho$.

\begin{figure}[!htb]
  \centering
\unitlength=2.00mm \special{em:linewidth 0.4pt}
\linethickness{0.4pt}
\begin{picture}(47.00,15.00)

\put(13.00,7.00){\oval( 8,14)}

\put(13.00 ,6.85){\line(1,0){6.30}}
 \put(13.00,7.15){\line(1,0){6.30}}
 \put(13.00 ,6.90){\line(0,1){7.00}}
\put(19.40 ,7.00){\vector(1,0){0.50}}

\put(14.50 ,8.50){\line(1,0){0.70}}
 \put(15.70,8.50){\line(1,0){0.70}}
  \put(17.00,8.50){\line(1,0){0.70}}
 \put(18.20 ,8.50){\vector(1,0){2.00}}

\put(14.50 ,9.50){\line(1,0){0.70}}
\put(15.70,9.50){\line(1,0){0.70}}
 \put(17.00,9.50){\line(1,0){0.70}}
\put(18.20 ,9.50){\vector(1,0){2.00}}

\put(14.50 ,10.50){\line(1,0){0.70}}
 \put(15.70,10.50){\line(1,0){0.70}}
  \put(17.00,10.50){\line(1,0){0.70}}
\put(18.20 ,10.50){\vector(1,0){2.00}}

\put(14.50 ,5.50){\line(1,0){0.70}}
 \put(15.70,5.50){\line(1,0){0.70}}
  \put(17.00,5.50){\line(1,0){0.70}}
  \put(18.20 ,5.50){\vector(1,0){2.00}}

\put(14.50 ,4.50){\line(1,0){0.70}}
 \put(15.70,4.50){\line(1,0){0.70}}
  \put(17.00,4.50){\line(1,0){0.70}}
 \put(18.20 ,4.50){\vector(1,0){2.00}}

\put(14.50 ,3.50){\line(1,0){0.70}}
 \put(15.70,3.50){\line(1,0){0.70}}
  \put(17.00,3.50){\line(1,0){0.70}}
\put(18.20 ,3.50){\vector(1,0){2.00}}

\put(11.50,10.00){\makebox(0,0)[cc]{$\varrho_m$}}
\put(11.50,7.00){\makebox(0,0)[cc]{$p$}}
\put(19.00,2.00){\makebox(0,0)[cc]{$\omega$}}

\put(31.00,7.00){\oval( 10,4)}

\put(26.00,7.00){\vector(-1,0){2.00}}
\put(31.00,10.50){\vector(0,-1){1.50}}
\put(31.00,3.50){\vector(0,1){1.50}}

\put(31.00,7.00){\makebox(0,0)[cc]{$\sigma_\perp$}}
\put(25.00,5.00){\makebox(0,0)[cc]{$e$}}

\end{picture}
\caption{Scattering of equivalent photons, forming the ``disk"
with radius $\varrho_m$, on the electron beam with radius
$\sigma_\perp$. }
 \label{fig:3}
\end{figure}
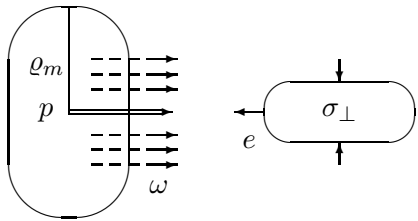

In the reference frame connected with the collider, the equivalent
photon with energy $\hbar \omega$ and the electron with energy
$E_e \gg \hbar \omega$ move toward each other (Fig.~\ref{fig:3})
and perform a Compton scattering. The characteristics of this
process are well known. The main contribution to the Compton
scattering is given by the region where the scattered photons fly
in a direction opposite to that of the initial photons. For such a
backward scattering, the energy of the equivalent photon $\hbar
\omega$ and the energy of the final photon $E_\gamma$ and its
emission angle $\theta_\gamma$ are related by
 \begin{equation}
    \hbar \omega = {E_\gamma \over 4 \gamma^2_e (1 - E_\gamma/E_e )}
   \left[1+ (\gamma_e\theta_\gamma)^2 \right]
 \label{1.1a}
 \end{equation}
and, therefore,
\begin{equation}
 \hbar \omega \sim {E_\gamma \over 4 \gamma^2_e (1 - E_\gamma/E_e
   )}\,.
 \label{1.1}
 \end{equation}

As a result, we find the radius of the ``disk'' of equivalent
photons with the frequency $\omega$ (corresponding to a final
photon with energy $E_\gamma$) as follows:
 \begin{equation}
 \varrho_m = {\gamma_p c\over  \omega} \sim \lambda_e\; {4 \gamma_e
\gamma_p}\, {E_e - E_\gamma \over E_\gamma} \,,\;\;\;
\lambda_e={\hbar\over m_ec}=3.86\cdot 10^{-11}\; {\rm cm}\,.
 \label{size}
 \end{equation}
For the HERA collider with $E_p=820$ GeV and $E_e=28$ GeV one
obtains
 \begin{equation}
 \varrho_m \gtrsim 1\; {\rm cm \ \ \ for\ \ \ } E_\gamma \lesssim
0.2 \;{\rm GeV}\ .
 \label{1.3}
 \end{equation}
Equation (\ref{size}) is also valid for the $e^- e^+ \rightarrow
e^- e^+ \gamma$ process with replacement protons by positrons. For
the VEPP-4 collider it leads to
 \begin{equation}
 \varrho_m \gtrsim 1\; {\rm cm \ \ \ for\ \ \ } E_\gamma \lesssim
15\; \mbox{ MeV }\,,
  \label{1.4}
  \end{equation}
for the PEP-II and KEKB colliders we have
 \begin{equation}
 \varrho_m \gtrsim 1\; {\rm cm \ \ \ for\ \ \ } E_\gamma \lesssim
0.1\; \mbox{ GeV }\,.
  \label{1.4a}
  \end{equation}

The  standard  calculation corresponds to the interaction of the
photons forming the ``disk'' with the unbounded flux of electrons.
However, the particle beams at the HERA collider have finite
transverse beam sizes of the order of $\sigma_\perp\sim 10^{-2}$
cm. Therefore, the equivalent photons from the region
$\sigma_\perp \lesssim \varrho \lesssim \varrho_m$ cannot interact
with the electrons from the other beam. This leads to the
decreasing number of the observed photons and the  ``observed
cross section''  $d \sigma_{\rm obs}$ is smaller than the
standard  cross section $d \sigma$ calculated for an infinite
transverse extension of the electron beam,
 \begin{equation}
   d \sigma_{\rm obs} = d \sigma - d \sigma_{\rm cor}.
  \label{1.5}
  \end{equation}
Here the correction $d \sigma_{\rm cor}$ can be presented in the
form
 \begin{equation}
 d \sigma_{\rm cor} = d \sigma_{\rm C}(\omega,\,E_\gamma) \
dn(\omega)
  \label{1.6}
  \end{equation}
where $dn(\omega)$ denotes the number of ``missing'' equivalent
photons and $d \sigma_{\rm C}$ is the cross section of the Compton
scattering.  Let us stress that the equivalent photon
approximation in this region has a high accuracy (the neglected
terms are of the order of $1/\gamma_p$). But for the qualitative
description it is sufficient to use the logarithmic approximation
in which this number is (see\cite{BLP}, \S 99)
 \begin{equation}
  dn = {\alpha \over \pi} {d\omega \over \omega} {d q_{\perp}^2 \over
q_{\perp}^2} \,.
 \label{1.7}
 \end{equation}
Since $q_\perp  \sim 1 / \varrho$, we can present the number of
``missing'' equivalent photons in the form
 \begin{equation}
 dn = {\alpha \over \pi} {d \omega \over \omega} { d\varrho^2 \over
\varrho^2}
 \label{1.8}
 \end{equation}
with the integration region in $\varrho$:
 \begin{equation}
 \sigma_\perp \lesssim \varrho \lesssim \varrho_m = {\gamma_p c
\over \omega}\,.
 \label{1.9}
 \end{equation}
As a result, this number is equal to
 \begin{equation}
 dn(\omega) = 2 {\alpha \over \pi} { d\omega \over \omega} \ln
{\varrho_m \over \sigma_\perp }\,,
 \label{1.10}
 \end{equation}
and the correction to the standard cross section with logarithmic
accuracy is\footnote{Within this approximation, the standard cross
section has the form
 $$
d\sigma = d \sigma_{\rm C} {\alpha \over \pi} {d\omega \over
\omega} {d q_{\perp}^2 \over q_{\perp}^2}={16\over 3} \alpha
r^2_e\, {dy\over y}\, \left(1-y+\mbox{${3\over 4}$} y^2\right)
\ln{4\gamma_e \gamma_p (1-y)\over y }
 $$
with the integration region $\hbar \omega /(c \gamma_p) \lesssim
\hbar q_\perp \lesssim m_e c$ corresponding to the impact
parameters $\varrho$ in the interval $\lambda_e \lesssim \varrho
\lesssim \varrho_{m}$. }
 \begin{equation}
  d\sigma_{\rm cor} = {16\over 3} \alpha r^2_e\, {dy\over y}\,
\left(1-y+\mbox{${3\over 4}$} y^2\right) \ln{4\gamma_e \gamma_p
(1-y)\lambda_e\over y \sigma_\perp}\,, \;\;y={E_\gamma\over
E_e}\,.
  \label{1.11}
 \end{equation}

\section{APPROXIMATIONS}

For future linear  $e^+ e^-$ colliders the transverse sizes of the
beams will change significantly during the time of interaction due
to a mutual attraction of very dense beams. However, for most of
the ordinary accelerators, including practically all $e^+e^-$ and
$ep$ storage rings, the change of the transverse beam sizes during
the collisions can be neglected. Below we use two main
approximations: 1) the particle movement in the bunches has a
quasi-classical character; 2) the particle distribution remains
practically unchanged during the collision.  Besides, in
calculating the coherent contribution, we neglect correlations in
particle coordinates.  All these approximation just the same as in
Ref.~\cite{BK02}. For definiteness, we use again the $ep$
collisions as an example.

Therefore, if the proton (electron) bunch moves along (opposite)
the direction of $z$-axis with the velocity $v_p$ ($v_e$), its
density has the form
 \begin{equation}
 n_p=n_p ({\mbox{\boldmath $\varrho$}}, z-v_pt)\,,\;\;
n_e=n_e ({\mbox{\boldmath $\varrho$}}, z+v_et)\,.
 \label{15}
 \end{equation}
We also introduce so called ``transverse densities''
 \begin{equation}
 n_p ({\mbox{\boldmath $\varrho$}})= \int n_p\,dz\,,\;\;
n_e ({\mbox{\boldmath $\varrho$}})= \int n_e\,dz
 \label{16}
 \end{equation}
which is equal to the total number of protons (electrons) which
cross a unit area around the impact parameter ${\mbox{\boldmath
$\varrho$}}$ during the collision.

Below we consider in detail the case when an electron deflection
angle $\theta_e$ is smaller than the typical radiation angle $\sim
1/\gamma_e$. It is easy to estimate the ratio of these angles. The
electric {\bf E} and magnetic {\bf B} fields of the proton bunch
are approximately equal in magnitude, $|{\bf E}| \approx  |{\bf
B}| \sim eN_p /(\sigma_x+\sigma_y)$. These fields are transverse
and they deflect the electron into the same direction. In such
fields the electron moves around a circumference of  radius $R\sim
\gamma_e m_e c^2/(eB)$ and gets the deflection angle $\theta_e
\sim l/R$. Therefore, the ratio of these angles is of the order of
\begin{equation}
{ \theta_e \over (1/\gamma_e)} \sim \eta={r_e N_p\over
\sigma_x+\sigma_y}  \,.
 \label{1}
\end{equation}
The parameter $\eta \gg 1$ only for the SLC and future linear
$e^+e^-$ colliders, in most of the colliders $\eta \lesssim 1$.

In our consideration we use the equivalent photon approximation.
In the region of interest (where impact parameters are large,
$\varrho \gtrsim \sigma_\perp$) this simple and transparent method
has a high accuracy. On the other hand, the operator
quasi-classical method, used in Ref.~\cite{BK02}, just coincides
in this region with the equivalent photon approximation.

\section{COHERENT AND INCOHERENT CONTRIBUTIONS}

\subsection{General formulae}

The corresponding formulae for the number of events in a single
collision of the electron and proton bunches can be found in
papers~\cite{Ginzyaf}, \cite{ESS1}. To calculate the MD-effect, we
need to know the distribution of equivalent photons (EP) for large
values of impact parameters. In this region we can consider the
electron--proton scattering as the scattering of electrons on the
electromagnetic field of the proton bunch. Replacing this field by
the flux of EP with some energy distribution, we obtain the number
of the produced photons in the form
\begin{equation}
dN_\gamma = dL_{\gamma e }(\omega ) \  d\sigma_{\rm C} (\omega,\,
E_\gamma ); \;\;\; dL_{\gamma e}(\omega )= n_\gamma
({\mbox{\boldmath $\varrho$}}, \omega) d\omega \,n_e
({\mbox{\boldmath $\varrho$}}) \, d^2\varrho \,.
 \label{18}
\end{equation}
Here $n_e ({\mbox{\boldmath $\varrho$}})$ is the transverse
electron density and $n_\gamma ({\mbox{\boldmath
$\varrho$}},\omega) d\omega$ is  the transverse density of EP with
the frequencies in the interval from $\omega$ to $\omega + d\omega
$. The quantity $dL_{\gamma e}(\omega )$ denotes the differential
luminosity for the collisions of EP and electrons and $d
\sigma_{\rm C} (\omega, E_\gamma)$ is the Compton cross section
for the scattering of the equivalent photon with the frequency
$\omega$ on the electron.

The transverse density of the EP is
 \begin{equation}
n_\gamma ({\mbox{\boldmath$\varrho$}},
 \omega )\;d\omega ={c\over 4\pi^2}\;\left\langle \mid {\bf E}_\omega
({\mbox{\boldmath $\varrho$}})\mid ^2\right\rangle\;{d\omega \over
\hbar \omega}
 \label{18a}
 \end{equation}
where ${\bf E}_\omega ({\mbox{\boldmath $\varrho$}})$ is the
spectral component of the collective electric field of the proton
bunch. The sign $\langle \dots \rangle$ denotes the averaging over
fluctuations of the field connected, for example, with the
fluctuations of particle positions for many collisions of bunches
in a given experiment. This field depends on a distribution of
charges in the proton bunch at $t=0$. We introduce the exact
(fluctuating) density of the proton bunch $n({\bf r})$ and the
averaging density
 \begin{equation}
  n_p({\bf r})= \langle n({\bf r}) \rangle
 \end{equation}
as well as their form factors
\begin{equation}
F({\bf q})= \int n({\bf r})\;\mbox {e}^{-{\rm i}{\bf qr}}\;
d^3r\,, \;\; F_p({\bf q})=    \langle F({\bf q}) \rangle=  \int
n_p({\bf r})\;\mbox {e}^{-{\rm i}{\bf qr}}\; d^3r
\end{equation}
with the normalization
\begin{equation}
F(0)=\int n({\bf r})\, d^3r = F_p(0)=N_p \ .
\end{equation}
In the classical limit
 \begin{equation}
  n({\bf r})=\sum_{a}\,\delta({\bf r}- {\bf r}_a)\,,\;\;
 F({\bf q})=\sum_{a}\,\mbox {e}^{-i{\bf q}{\bf r}_a}
 \label{22a}
 \end{equation}
where ${\bf r}_a$ is the radius-vector of the $a$-th proton. In
these notations, the exact (fluctuating) collective field is
 \begin{equation}
 {\bf E}_\omega ({\mbox{\boldmath $\varrho$}})=- { {\rm i} e \over
\pi c} \ \int{{\bf q}_\bot \; \mbox {e}^{{\rm i}{\bf q}_\bot
{\mbox{\scriptsize\boldmath $\varrho$}}}\over {\bf
 q}^2_\bot+\omega^2/(c\gamma_p)^2} \,F({\bf q})
 \; d^2 q_\bot\,
 \label{12b}
\end{equation}
with $q_z= \omega/c$.

As a result, the number of events
 \begin{equation}
 dN_\gamma  \propto n_\gamma ({\mbox{\boldmath$\varrho$}},
 \omega ) ={\alpha\over 4\pi^4 \omega}
\int\, {({\bf q}_\bot {\bf q}'_\bot) \; \mbox{e}^{{\rm i}({\bf
q}_\bot -{\bf q}'_\bot) \mbox{\scriptsize\boldmath $\varrho$}}
\over  [{\bf q}^2_\bot+\omega^2/(c\gamma_p)^2]
 [({\bf q}'_\bot)^2+\omega^2/(c\gamma_p)^2]}
\;A\; d^2 q_\bot\,d^2 q'_\bot\;
 \end{equation}
depends on the quantity
 \begin{equation}
 A  =\left\langle \, F({\bf q})\,F^*({\bf q}')\, \right\rangle =
\int \left\langle \,n({\bf r})\, n({\bf r}')\,\right\rangle\;\mbox
{e}^{-{\rm i}({\bf qr} -{\bf q}'{\bf r}')}\,d^3r\, d^3r'
 \label{sum}
 \end{equation}
in which
 \begin{equation}
  q_z=q'_z= \omega/c\,.
 \label{qo}
 \end{equation}

\subsection{Coherent and incoherent bremsstrahlung}

The obtained general formulae include the  coherent and incoherent
contributions. The coherent contribution is determined by the
average field which is given  by Eq. (\ref{12b}) with the
replacement $F({\bf q})$ by $F_p({\bf q})$ or with the replacement
of the exact density by the average density. The averaged density
of the proton bunch has a single scale in the longitudinal
direction --- the length of the bunch $l$. Therefore, the average
field of the bunch is essential in the region of frequencies
$\omega =cq_z\lesssim c/l$ and should be small in the region of
large frequencies $\omega \gg c/l$. In particular, if the proton
bunch has the Gaussian distribution, its form factor is equal to
\begin{equation}
F_p({\bf q})=N_p\; \mbox{exp} \left[ - \mbox{${1\over
2}$}(q_x\sigma_{x})^2 - \mbox{${1\over 2}$} (q_y\sigma_{y})^2 -
 \mbox{${1\over 2}$}
(\omega l/c)^2 \right]
 \label{G}
\end{equation}
and vanishes in the discussed region of frequencies from the
interval $c/l \ll \omega \ll c/a$ where $a$ is the mean distance
between particles.

A bunch at colliders  can be treated as a continues media with a
smooth average particle distribution of the Caussian type. If we
neglect the correlations of the particle coordinates in such
media, the average product of densities $\left\langle n({\bf
r})\,n({\bf r}')\right\rangle$ is expressed only via the average
densities as follows (see, for example,~\cite{LL5})
\begin{equation}
\left\langle n({\bf r})\,n({\bf r}')\right\rangle={n}_p({\bf
r})\,{n}_p({\bf r}')+ \delta({\bf r}- {\bf r}')\,{n}_p({\bf r})\,.
 \label{sigma}
 \end{equation}
As a result, the quantity $A$ can be presented as a sum of
coherent and incoherent contributions:
 \begin{equation}
 A =A_{\rm coh}+  A_{\rm incoh}\,.
 \end{equation}

The coherent contribution is related to the first item ${n}_p({\bf
r})\,{n}_p({\bf r}')$  in Eq.~(\ref{sigma}) and is equal to
 \begin{equation}
  A_{\rm coh}= F_p({\bf q})\,F_p^*({\bf q}')\,.
 \label{COH}
  \end{equation}
This formula was used in Refs.~\cite{Ginzyaf}, \cite{ESS1} to
obtain main characteristics of the coherent bremsstrahlung. It
also allows us to obtain the following estimate for the Gaussian
beams in the region of interest (at $|q_x| \lesssim 1/ \sigma_{x}$
and $|q_y| \lesssim 1/ \sigma_{y}$):
 \begin{equation}
 A_{\rm coh}\sim N_p^2\,\mbox{exp} \left[ - (\omega l/c)^2
\right]\,.
 \end{equation}

The incoherent contribution is connected with the second item
$\delta({\bf r}- {\bf r}')\,{n}_p({\bf r})$ in Eq.~(\ref{sigma})
and is equal to (taking into account Eq. (\ref{qo}))
 \begin{equation}
 A_{\rm incoh}= F({\bf q}_\bot -{\bf q}'_\bot)\,.
 \label{INC}
 \end{equation}
Note, that this expression is determined by the transverse average
density of the proton bunch and it does not depend on $\omega$.
For the Gaussian beams in the region of interest, we obtain from
(\ref{INC}) an estimate
 \begin{equation}
 A_{\rm incoh} \sim N_p\,.
 \end{equation}
Formula (\ref{INC}) was used to derive the previous results about
MD-effect (for details see review~\cite{KSS}). It is seen from the
above consideration that the incoherent contribution for usual
colliding beams has no ``additional subtraction'' related to the
average electromagnetic field of the proton bunch.

\subsection{Comparison with the approach used in~\cite{BK02}}

We derive the final expression for the incoherent contribution
from general equations (\ref{18}), (\ref{18a}) and (\ref{12b}) as
a simple consequence of natural assumptions about the particle
distribution in a proton bunch. It is useful to rewrite these
equations in the form convenient for comparison with the
corresponding equations in~\cite{BK02}. To do this, we note that
the Compton cross section $d\sigma_{\rm C} (\omega,
E_\gamma)\propto |{\bf e}{\bf M}_{\rm C} |^2$ where ${\bf e}{\bf
M}_{\rm C}$ is the amplitude of the Compton scattering for the EP
with the polarization vector ${\bf e}$. Therefore, the number of
events
 \begin{equation}
 dN_\gamma\propto  |M|^2\,,\;\; M={\bf E}_\omega ({\mbox{\boldmath
$\varrho$}}){\bf M}_{\rm C}
 \end{equation}
where the quantity $M$ is proportional to the probability
amplitude of the process. Further, we use Eq. (\ref{22a}) and the
well known equality
 \begin{equation}
 \int  {{\bf q}_\perp \, \mbox{e}^{{\rm i}{\bf q}_\bot
\mbox{\scriptsize\boldmath $\varrho$}} \over {\bf q}^2_\perp
+(1/b)^{2}}\, d^2q_\perp = {2\pi {\rm i}\over b}\,{\mbox{\boldmath
$\varrho$} \over \varrho} \,K_1(\varrho/b)
 \end{equation}
where $K_n(x)$ denotes the modified Bessel function of third kind
with integer index $n$ (McDonald function). Then we obtain
 \begin{equation}
 M=\sum_{a=1}^{N_p} M_a\,,\;\; M_a= {\bf E}^{(a)}_\omega
({\mbox{\boldmath $\varrho$}})\,{\bf M}_{\rm C}\,,\;\; {\bf
E}^{(a)}_\omega ({\mbox{\boldmath $\varrho$}})=- {2e\over c
\varrho_m}\, {\mbox{\boldmath $\varrho$}'_a\over \varrho'_a}
K_1\left(\varrho'_a / \varrho_m\right)\, {\rm e}^{-{\rm i} \omega
z_a/c}
 \label{Ma}
 \end{equation}
where $\mbox{\boldmath $\varrho$}'_a=\mbox{\boldmath
$\varrho$}_a-\mbox{\boldmath $\varrho$}$ is the impact parameter
between the $a$-th proton and the electron and the parameter
$\varrho_m=\gamma_pc/\omega$ is the radius of the ``disc'' of EP
(see Fig.~3). The item $M_a$ is the contribution to $M$ related to
the interaction of the electron with the $a$-th proton.

Our expression for $M_a$ coincides with the corresponding
expression in~\cite{BK02} with the only (but essential!)
exception: in paper~\cite{BK02} the factor
 \begin{equation}
 {\rm e}^{-{\rm i} \omega z_a/c}
 \label{phase}
 \end{equation}
is absent. It means that in Ref.~\cite{BK02} it was neglected by
the dependence of the exact and average density of the proton
bunch on longitudinal coordinates. However, according to the
consideration given above, this dependence is crucial for
description of the coherent contribution.

Let us clarify this point by the following simple calculations.
Our incoherent contribution corresponds to the average square of
the individual fields of protons, i.e. to the sum
 \begin{equation}
S=\sum_a \, \langle\, |M_a|^2 \,\rangle
 \end{equation}
in which the expression $ |M_a|^2$ does not depend on $z_a$. In
that case the sum over $a$ transforms to the following integral
over the transverse coordinates
 \begin{equation}
 \sum_a\langle\, \ldots \rangle \to \int d^2 \varrho_a dz_a \, n_p(\mbox{\boldmath
$\varrho$}_a,z_a)\,\ldots=\int d^2\varrho_a \, n_p(\mbox{\boldmath
$\varrho$}_a)\,\ldots\,.
 \end{equation}
After the substitution $\mbox{\boldmath $\varrho$}_a =
\mbox{\boldmath $\varrho$}+{\bf r}_\perp$, we obtain the
expression
 \begin{equation}
S= {4e^2\over c^2 \varrho^2_m}\,|{\bf e}{\bf M}_{\rm C}|^2\, \int
n_p( \mbox{\boldmath $\varrho$}+{\bf r}_\perp)\,
K_1^2(r_\perp/\varrho_m) \, d^2 r_\perp
 \end{equation}
which is equivalent to the previous result (\ref{INC}).

It was claimed in paper~\cite{BK02} that an additional subtraction
$S_1$ has to be done which corresponds to the square of average
individual fields of protons, i.e.
 \begin{equation}
 S\to S-S_1\,,\;\; S_1= \sum_a \left|\langle\, M_a \,\rangle\,\right|^2\,.
 \end{equation}
As we can jude from the final expression for $S_1$, the averaging
in this case means averaging over the transverse coordinates of
protons only,
 \begin{equation}
 \langle\, M_a \,\rangle= \int  M_a \,{n_p(\mbox{\boldmath
$\varrho$}_a)\over N_p} \, d^2\varrho_a\,.
 \label{Ma1}
 \end{equation}
It would be natural if the expression $M_a$ would not depend on
the longitudinal coordinates of protons $z_a$.

But according to our analysis, it is not the case since
 \begin{equation}
 M_a \propto  {\rm e}^{-{\rm i} \omega z_a/c}\,.
 \end{equation}
Taking into account this very fact, one has to perform averaging
over the longitudinal coordinates $z_a$ as well, i.e. instead of
(\ref{Ma1}) we have
 \begin{equation}
 \langle\, M_a \,\rangle= \int  M_a \, {n_p(\mbox{\boldmath
$\varrho$}_a, z_a )\over N_p} \, d^2\varrho_adz_a\,.
 \end{equation}
It changes the final result dramatically, since in the discussed
region of frequencies $\omega \gg c/l$ the expression $\langle\,
M_a \,\rangle$ disappears. For example, in the case of the
Caussian distribution we have
 \begin{equation}
 \langle\, M_a \,\rangle\propto \mbox{exp} \left[- \mbox{${1\over
2}$} (\omega l/c)^2 \right]\,\ll\, 1 \;\;\mbox{at } \;\;\omega \gg
c/l\,.
 \end{equation}

\section{CONCLUSIONS}

Let us compare the coherent and incoherent contributions for the
Gaussian beams. In this case, the ratio
 \begin{equation}
 {A_{\rm coh}\over A_{\rm incoh}} \sim
 N_p\,\mbox{exp} \left[ - (\omega l/c)^2
\right]
 \end{equation}
is determined by the parameters $\omega l/c$. Since $ \hbar \omega
\sim E_\gamma /[4 \gamma_e^2 (1-E_\gamma/E_e)]$, it is also useful
to introduced the coherence length
 \begin{equation}
  l_{\rm coh} = {4 \gamma_e^2 \hbar c\over
 E_\gamma}\,(1-E_\gamma/E_e)
 \end{equation}
and the critical energy for the coherent bremsstrahlung
 \begin{equation}
  E_c ={4 \gamma_e^2 \hbar c\over l}\,.
 \end{equation}

If the coherence length is large, $l_{\rm coh}\gtrsim l$, or if
the final photon energy is small, $E_\gamma \lesssim E_c$, the
parameter $\omega l/c \lesssim 1$ and the coherent contribution is
dominant.

On the contrary, in the region of large photon energy, $E_\gamma
\gg E_c$, or small coherence length, $l_{\rm coh}\ll l$,
considered in paper~\cite{BK02}, the incoherent contribution
dominates. In particular, at $\omega l/c > 6$ and $N_p \sim
10^{11}$, the ratio
 \begin{equation}
 {A_{\rm coh}\over A_{\rm incoh}} \sim
 N_p\,\mbox{e}^{ - 36} \ll 1\,,
 \end{equation}
the coherent contribution is completely negligible and the
previous formulae for the MD-effect are valid.

This consideration shows that the effect, derived in
paper~\cite{BK02}, is absent just in the region, discussed in this
paper. At the end of this section we reconsider the experiments
analyzed in paper~\cite{BK02}.

The HERA experiment~\cite{Piot95}. In this case $E_e= 27.5$ GeV
and $l= 8.5$ cm, therefore, $E_c = 27$ keV. For the observed
photon energies $E_\gamma =2\div 8$ GeV, the parameter
 \begin{equation}
 {\omega l\over c} \sim {E_\gamma\over E_c} > 10^{4}\,,
 \end{equation}
and the coherent contribution is completely
negligible\footnote{Moreover, in the HERA experiment the coherence
length is of the order of the mean distance between particles in
the proton bunch, $l_{\rm coh}\sim a\sim \left(l\sigma_x
\sigma_y/N_p\right)^{1/3}$.}. Therefore, the new correction to the
previous results on the level of $10$ \%, obtained in~\cite{BK02},
is, in fact, absent.

The VEPP-4 experiment~\cite{Blinov82}. In this case $E_e= 1.84$
GeV and $l= 3$ cm, therefore, $E_c = 0.34$ keV. For the observed
photon energies $E_\gamma \gtrsim 1$ MeV, the parameter
 \begin{equation}
 {\omega l\over c} \sim {E_\gamma\over E_c} > 10^{3}\,,
 \end{equation}
and the coherent contribution is completely negligible.

The case of a ``typical linear collider'' with $E_e=500$ GeV and
$E_\gamma = E_e/1000$. This example,  considered in
paper~\cite{BK02}, is irrelevant for the discussed problem, since
the coherent radiation (or beamstrahlung) at a typical linear
collider is absolutely dominated in this very region over the
ordinary incoherent bremsstrahlung --- see, for example, the TESLA
project~\cite{TDR}.

\begin{acknowledgments}

We are very grateful to V.~Dmitriev, V.~Fadin, I.~Ginzburg,
V.~Katkov, I.~Kolokolov and A.~Milshtein for useful discussions.

\end{acknowledgments}

\end{document}